\RequirePackage[2020-02-02]{latexrelease}
\documentclass[prl,twocolumn,superscriptaddress,preprintnumbers,amsmath,amssymb]{revtex4}

\usepackage{amsmath,bm}
\usepackage{graphicx}
\usepackage{dcolumn}
\usepackage{color}
\DeclareGraphicsExtensions{.png .jpg .pdf .eps}
\usepackage{float}
\usepackage{mathrsfs}
\usepackage{isotope}
\usepackage{xfrac} 
\usepackage{physics} 
\usepackage{xcolor}
\usepackage{comment}

\begin{document}

\title{Striped excitonic (super)solid in anisotropic semiconductors with screened exciton interactions}

\author{J. F. de Oliveira Neto}\email{fernandes.neto@fisica.ufc.br}
\author{F. M. A. Guimarães}\email{mateus.guimaraes@fisica.ufc.br }
\affiliation{Departamento de F\'isica, Universidade Federal do Cear\'a, Caixa Postal 6030, Campus do Pici, 60455-900 Fortaleza, Cear\'a, Brazil}

\author{Davi S. Dantas}\email{davi.dantas@ifce.edu.br }
\affiliation{Departamento de F\'isica e Matem\'atica, Instituto Federal de Educa\c{c}\~ao, Ci\^encia e Tecnologia do Cear\'a - IFCE, Av. Treze de Maio 2081, 60040-215 Fortaleza, Cear\'a, Brazil}

\author{F. M. Peeters}\email{francois.peeters@uantwerpen.be}
\affiliation{Departamento de F\'isica, Universidade Federal do Cear\'a, Caixa Postal 6030, Campus do Pici, 60455-900 Fortaleza, Cear\'a, Brazil}
\affiliation{Department of Physics, University of Antwerp, Groenenborgerlaan 171, B-2020, Antwerp, Belgium}
\affiliation{Nanjing University of Science and Technology, Nanjing 210044, China}

\author{M. V. Milo\v{s}evi\'{c}}\email{milorad.milosevic@uantwerpen.be}
\affiliation{Department of Physics, University of Antwerp, Groenenborgerlaan 171, B-2020, Antwerp, Belgium}
\affiliation{NANOlight Center of Excellence, University of Antwerp, Belgium}
\affiliation{Instituto de Física, Universidade Federal de Mato Grosso, Cuiabá, Mato Grosso 78060-900, Brazil}

\author{A. Chaves}\email{andrey@fisica.ufc.br}
\affiliation{Departamento de F\'isica, Universidade Federal do Cear\'a, Caixa Postal 6030, Campus do Pici, 60455-900 Fortaleza, Cear\'a, Brazil}
\affiliation{Department of Physics, University of Antwerp, Groenenborgerlaan 171, B-2020, Antwerp, Belgium}


\begin{abstract}

Within the Gross-Pitaevskii framework, we reveal the emergence of a crystallized phase of an exciton condensate in an atomically-thin anisotropic semiconductor, where screening of exciton-exciton interactions is introduced by a proximal doped graphene layer. While such screened interactions are expected to yield a hexagonal crystal lattice in the excitonic condensate in isotropic semiconductor quantum wells [see e.g. Phys. Rev. Lett. \textbf{108}, 060401 (2012)], here we show that for atomically thin semiconductors with strong electronic anisotropy, such as few-layer black phosphorus, the crystallized exciton phase acquires a parallel stripe structure - unanticipated to date. The optimal conditions for the emergence of this phase, as well as for its coexistence with excitonic superfluidity in a striped supersolid phase, are identified.

\end{abstract}

\maketitle

\textit{Introduction} - A supersolid is a unique state of matter that combines the long-range order of a solid with the frictionless flow characteristic of superfluids. Although initially predicted in solid helium several decades ago \cite{andreev1969quantum,chester1970speculations,leggett1970can}, its experimental confirmation has only recently been achieved. Supersolidity was first observed in quasi-one-dimensional (quasi-1D) Bose-Einstein condensates (BECs) of alkaline atoms coupled to external light \cite{li2017stripe,leonard2017supersolid,leonard2017monitoring}, followed by similar discoveries in 1D and two-dimensional (2D) lanthanide atomic BECs \cite{bottcher2019transient,tanzi2019observation,chomaz2019long,norcia2021two}. Despite these advances, a definitive evidence of supersolidity in helium remains elusive \cite{kim2012absence}, prompting investigations into whether this phase might also be realized in other platforms.

Among the various candidates to enable supersolidity, electron-hole pairs (excitons) have emerged as particularly promising for realizing the supersolid phase \cite{parish2011supersolidity,zarenia2017inhomogeneous,conti2023chester,zeng2023evidence,muszynski2024observation}. In semiconductors, excitons behave as composite bosons and, as such, are expected to undergo BEC at low temperatures. Indeed, signatures of exciton condensation have been observed in double quantum well structures, where an external bias across the wells spatially separates electrons and holes, thereby prolonging the exciton lifetime \cite{butov2002towards}, as well as in a gas of 1s para-exciton states in bulk CuO$_2$ \cite{morita2022observation}. Recent experiments \cite{wang2019evidence, chaves2019exotic} have provided further evidence for exciton condensation in systems comprising monolayers of WSe$_2$ and MoSe$_2$, which serve as quantum wells separated by a few-layer hBN spacer. In these heterostructures, the inherent band offset between the materials ensures vertical electron-hole separation, leading to extended exciton lifetimes even in the absence of an external bias. More recently, superfluidity has also been proposed as a potential phase in an experimentally observed quantum critical point in few-layer black phosphorus (BP) \cite{saberi2018high,zheng2022quantum,huang2024bright,milovsevic2024bright}. The anisotropic crystal structure of BP introduces additional complexity, resulting in excitons with anisotropic effective masses in the material plane \cite{saberi2018anisotropic}, thus offering a unique platform to explore the properties of BECs with anisotropic mass.

Building on these developments, in this Letter, we theoretically investigate a system where excitons are confined within a few-layer anisotropic mass semiconductor (as a practical example, we have chosen 3-layer BP \cite{castellanos2015black, castellanos2014isolation,ling2015renaissance,xia2014rediscovering}) spatially separated by a few-layer hBN spacer from a proximate 2D electron gas (2DEG), formed by electron-doped bilayer graphene. A vertical bias across the system induces a dipole-like behavior in the excitons, i.e.  spatially separates the electron and the hole.

As highlighted in Refs. \cite{shelykh2010rotons,matuszewski2012exciton}, the presence of the 2DEG screens the dipole-dipole interaction between excitons, yielding a partially attractive and partially repulsive interaction, deviating from the purely repulsive interaction typically observed in exciton experiments. This scenario is analogous to the behavior found in lanthanide BECs, where the attractive dipole-dipole interaction, counterbalanced by contact interactions, kinetic energy and quantum fluctuations, gives rise to the formation of the supersolid state. Indeed, by carefully balancing the densities of electrons and excitons, and precisely tuning the semiconductor-graphene separation, our calculation predicts the emergence of a supersolid phase with striking similarities to those observed in atomic BECs. However, unlike in the lanthanide atomic case, where quantum fluctuations are essential for stabilizing the supersolid phase \cite{wachtler2016quantum,bisset2016ground}, our exciton-based system achieves stability purely through the balance between the dipole-dipole interaction, contact interaction, and kinetic energy. 

Our results further predict that the hexagonal, Wigner-like crystal, expected in BECs for systems based on isotropic mass excitons in semiconductor quantum wells \cite{matuszewski2012exciton, conti2023chester}, transits into a pattern of parallel stripes when dealing with excitons possessing anisotropic masses in BP. This striped phase represents a novel configuration unique to anisotropic mass systems. Moreover, we demonstrate that these stripes can coexist with superfluidity, forming a striped supersolid phase reminiscent of that experimentally found in Ref. \cite{li2017stripe} for ultracold alkaline gases, where the stripes were imposed by an external optical trap. In contrast, in our system, these stripes arise spontaneously, without any externally imposed pattern, showing striking similarities to those theoretically predicted in lanthanide BECs, where the stripes emerge due to the anisotropy of the dipole-dipole interaction \cite{aleksandrova2024density} in a high-density atoms limit \cite{zhang2021phases,hertkorn2021pattern,schmidt2022self}. However, due to the experimental limitations in atom density, these states in lanthanide BECs remain beyond the reach of current experiments. Our phase diagram reveals an accessible parameter range for the striped supersolid phase, further underscoring the richness and complexity of the exciton-based supersolid phase. 

\begin{figure}[]
\centering
\includegraphics[width=1 \linewidth]{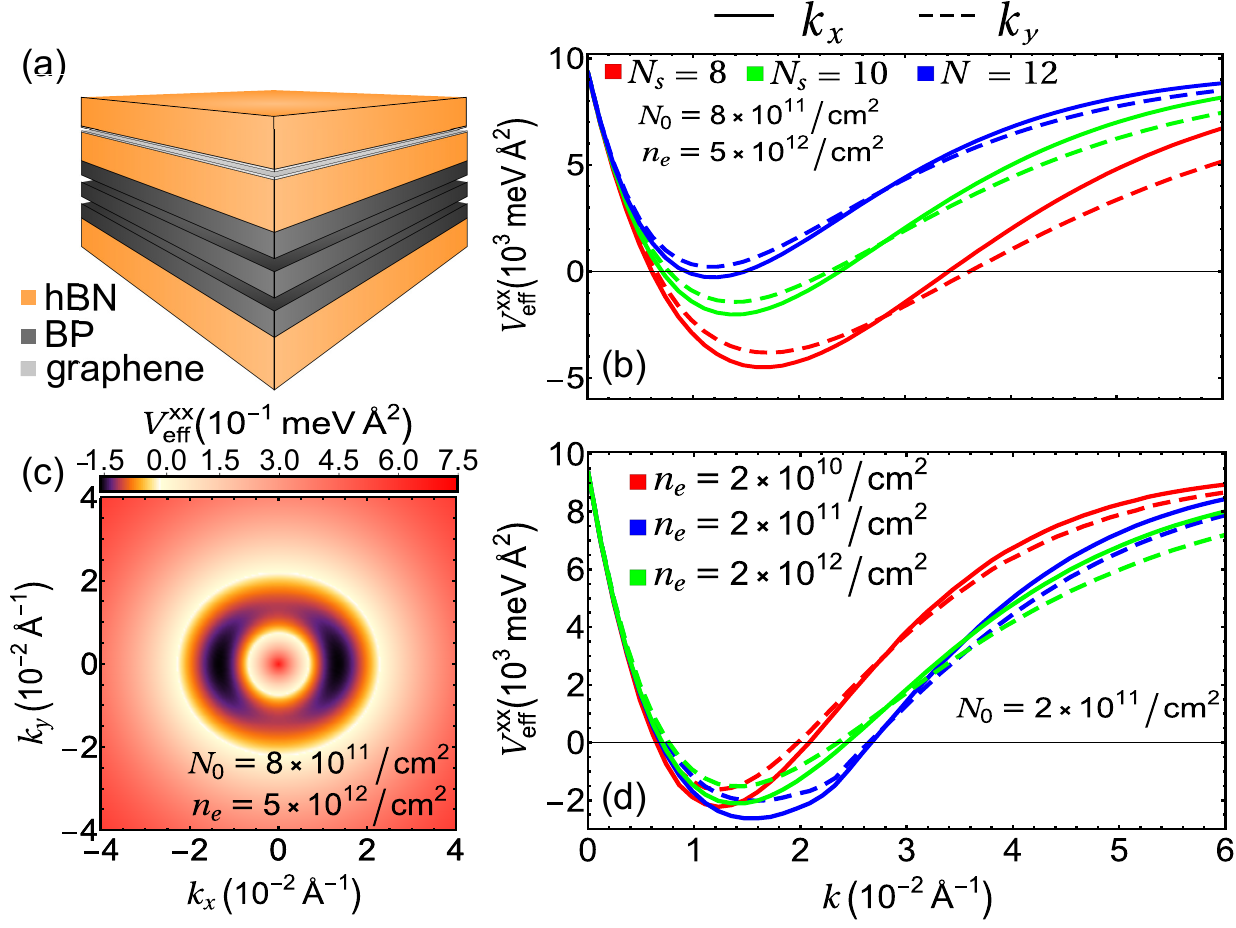}
\caption{(color online) (a) Sketch of the system, consisting of a 3-layer BP slab separated by a $N_{s}$-layer hBN spacer from a 2DEG in a bilayer graphene. The whole system is encapsulated in hBN. (b) Screened exciton-exciton interaction potential in reciprocal space  for three values of spacer thickness $N_s$. Results along the $k_{x(y)}$-direction are represented by solid (dashed) curves. (c) A color map of this potential for $N_s = 10$, $n_e = 5 \times 10^{12}$cm$^{-2}$, and $N_0 = 8 \times 10^{11}$cm$^{-2}$. (d) The same as in (b), but for three values of graphene doping $n_e$.}
\label{fig.sketch}
\end{figure}

\textit{Theoretical framework} - We consider a hybrid Bose-Fermi system comprising a 2DEG in bilayer graphene and an exciton gas in a few-layer 2D semiconductor with anisotropic mass, specifically a 3-layer BP structure, encapsulated by \isotope[]{hBN}{}. The semiconductor layers are spatially separated from the 2DEG by a $N_{s}$-layer \isotope[]{hBN}{} spacer, with a thickness of $d = N_{s} \times 3.32$ Å, as illustrated in Fig. \ref{fig.sketch}(a). 

At zero temperature, the dynamics of the exciton BEC is described by the time-dependent Gross-Pitaevskii equation (GPE),
\begin{equation} \label{GPE}
  i \hbar \pdv{}{t}\Psi(\mathbf{r},t)= \left[-\frac{\hbar^2}{2\,m^{\text{x}}_x}\frac{\partial^2}{\partial x^2}-\frac{\hbar^2}{2\,m^{\text{x}}_y}\frac{\partial^2}{\partial y^2} + \Phi(\mathbf{r},t) \right]\Psi(\mathbf{r},t),
\end{equation}
which accounts for the mean-field behavior of the condensate wave function $\Psi(\mathbf{r},t)$ in the response of anisotropic effective mass $m^{\text{x}}_{x (y)}$ in the $x(y)$ direction and exciton-exciton interactions, captured by $\Phi(\mathbf{r},t)=\int  V^{\text{xx}}_{\text{eff}}(\mathbf{r} - \mathbf{r}',t)\abs{\Psi(\mathbf{r},t)}^2 \dd[]{ \mathbf{r}'}$. Here, the order parameter is normalized to the total number of excitons, while the potential $V^{\text{xx}}_{\text{eff}}(\mathbf{r} - \mathbf{r}',t)$ represents the effective exciton-exciton interaction screened by the 2DEG.

In reciprocal space, this potential is expressed as \cite{shelykh2010rotons,matuszewski2012exciton}
\begin{equation}\label{exciton_interaction}
    \begin{split}
        V^{\text{xx}}_{\text{eff}}(\mathbf{k},\omega)=\frac{\left[V_\mathbf{k}^{\text{xx}}+\frac{\left(V^{\text{ex}}_\mathbf{k}\right)^2\Pi_{\mathbf{k},\omega}}{1-V^{\text{ee}}_\mathbf{k}\Pi_{\mathbf{k},\omega}}\right]\left[(\hbar\omega)^2-\left(E^{\text{x}}_\mathbf{k}\right)^2\right]}{(\hbar\omega)^2-\left(E^{\text{x}}_\mathbf{k}\right)^2-2N_0\hspace{-.1cm}\left[\hspace{-.08cm}V^{\text{xx}}_{\mathbf{k}}+\frac{\left(V^{\text{ex}}_\mathbf{k}\right)^2\Pi_{\mathbf{k},\omega}}{1-V^{\text{ee}}_\mathbf{k}\Pi_{\mathbf{k},\omega}}\hspace{-.1cm}\right]\hspace{-.1cm}E^{\text{x}}_\mathbf{k}},
    \end{split}
\end{equation}
which couples the electron-electron (or hole-hole) interaction in the proximal graphene layer $(V^{\text{ee}}_\mathbf{k})$, the exciton-exciton interaction in the 3-layer BP $(V^{\text{xx}}_\mathbf{k})$, and the electron-exciton interaction potential $(V^{\text{ex}}_\mathbf{k})$. Moreover, the presence of the 2DEG directly influences the system's Bogoliubov dispersion, resulting in $\left(\hbar \omega\right)^2=\left(E^{\text{x}}_\mathbf{k}\right)^2+2N_0 V^{\text{xx}}_{eff}E^{\text{x}}_\mathbf{k}$. In Eq. \eqref{exciton_interaction}, $N_0$ represents the occupation number of the condensate, $E^{\text{x}}_\mathbf{k}$ is the exciton dispersion, and $\Pi_{\mathbf{k},\omega}$ is the polarization matrix due to the electron gas, given by
\begin{equation}
    \Pi(\mathbf{k},\omega) =\sum_{\mathbf{q}} \frac{f_{\mathbf{k}-\mathbf{q}}-f_{\mathbf{k}}}{\hbar\omega+i\hbar\delta+E^{e}_{\mathbf{k}-\mathbf{q}}-E^{e}_{\mathbf{k}}},
\end{equation}
where $f_{\mathbf{k}}$ and $E^{\text{e}}_\mathbf{k}$ are the Fermi distribution and the dispersion of the bare electron in the 2DEG, respectively. 

We emphasize that a similar potential was first proposed in Refs. \cite{shelykh2010rotons,matuszewski2012exciton} and has been adapted here to account for the anisotropic 2D semiconductor environment screened by 2DEG in the bilayer graphene. The electron dispersion $E^{e}_\mathbf{k}$ of the bilayer graphene is modelled by a quadratic dispersion, with an effective mass $m^e = 0.054 m_0$ \footnote{We have verified that assuming a monolayer graphene instead of bilayer graphene as 2DEG would yield the same BEC phases we discussed  here. This modification in the 2DEG material results simply in a linear dispersion for $E^{e}_k$ and, consequently, a different dependence of the Fermi energy on the electron density $n_e$, but final results remain qualitatively the same.}. Besides, the potential $V^{\text{ex}}_\mathbf{k}$ now incorporates the material's anisotropy. A detailed derivation of this potential is provided in the Appendix, resulting in 
    \begin{equation}\label{eq. ex2deg potential}
    \begin{split}
         V^{\text{ex}}_{\mathbf{k}}(\mathbf{k})=&\frac{e^2}{2\,\epsilon}\frac{e^{-k L}}{k}\left[V_+(\mathbf{k}) + V_-(\mathbf{k})\right],
    \end{split}  
\end{equation}
where
\begin{equation}
    V_{\pm}(\mathbf{k}) = \frac{e^{\pm\frac{d}{2}k}}{\left[1+\left(\frac{\beta^{e(h)}_x a_B k_x}{2\mu_{x}}\right)^2+\left(\frac{\beta^{e(h)}_y a_B k_y}{2\mu_{y}}\right)^2\right]^{3/2}}. 
\end{equation}
Here, $L$ is the distance between the graphene and BP, and the $-(+)$ sign denotes the electron (hole) contribution, with $\beta^{e(h)}_{x(y)}=m^{e(h)}_{x(y)}/(m^e_{x(y)}+m^h_{x(y)})$, where $m^{e(h)}_{x(y)}$ is the effective mass of electron (hole) in the $x(y)$-direction in BP. This expression is derived under the assumption that excitons exhibit a vertical dipole moment $ed$, which can be induced and controlled via a vertical bias \cite{chaves2021signatures}, and that the exciton relative motion wave function follows an anisotropic hydrogenic approximation, with a direction dependent effective Bohr radius $a_B/\mu_{x,y}$, where $a_B = a_0 \epsilon$, $a_0$ is the Bohr radius of the hydrogen atom, $\epsilon$ is the dielectric constant of the environment, taken as $\epsilon = 4.5 \epsilon_0$ for hBN, and $\mu_{x(y)} = (1/m^e_{x(y)} +1/m^h_{x(y)})^{-1}$ is the electron-hole reduced mass in the $x(y)$-direction. It is straightforward to verify that if one takes $m_x^{e,h}= m_y^{e,h}$, Eq. (\ref{eq. ex2deg potential}) reduces to the exciton-electron interaction potential assumed for isotropic systems given in previous papers \cite{matuszewski2012exciton,plyashechnik2023coupled}. 

The potentials $V^{\text{ee}}_{\mathbf{k}}$ and $V^{\text{xx}}_{\mathbf{k}}$ are assumed to be in the form of Coulomb and contact \cite{tassone1999exciton, lobanov2016theory, cotlect2016superconductivity} interactions, respectively \cite{shelykh2010rotons,matuszewski2012exciton}:
\begin{equation}\label{eq.simpleinteractions}
V^{\text{ee}}(k) = \frac{2\pi e^2}{k\,\epsilon}, \quad \quad\quad       
V^{\text{xx}} (k) = \frac{e^2d\, \xi}{\epsilon_{BP}},
\end{equation}
where $\xi = 0.05$ is a correlation correction \cite{matuszewski2012exciton}, relevant in 2D materials where electron and hole wave functions are not confined to a single layer and exhibit significant overlap. These approximations hold provided that (i) electrons and holes are split along the out-of-plane direction by an applied electric field, which helps enhancing exciton lifetime, (ii) the thickness of the proximal graphene layer is much smaller than the characteristic distances between excitons, as set by exciton density, and (iii) the hBN spacer is significantly thicker than the BP slab that hosts the excitons, as is the case here. Note that the anisotropy in the static dielectric constant of BP ($\epsilon_x = 12.5 \epsilon_0$, $\epsilon_y = 10 \epsilon_0$, $\epsilon_z = 8 \epsilon_0$) is not as pronounced as that in its effective mass. Therefore, in Eq. \eqref{eq.simpleinteractions}, a geometric mean of the dielectric constants, $\epsilon_{BP} = (\epsilon_x \epsilon_y \epsilon_z)^{1/3} \approx 10 \epsilon_0$, suffices. For more details on the validity of these assumptions and approximations, refer to the Appendix.

As an example of the influence of anisotropic mass, Fig. \ref{fig.sketch}(b) shows the resulting screened exciton-exciton potential in reciprocal space, assuming $n_e = 5 \times 10^{12}\;\text{cm}^{-2}$, $N_0 = 8 \times 10^{11}\;\text{cm}^{-2}$, and different values of the spacer thickness $N_s = 8, 10,$ and 12.  In all cases, the anisotropy in the energy dispersion and effective masses of BP results in an anisotropic interaction potential, as illustrated by the different curves for wave vectors pointing in the $k_x$ (solid lines) and $k_y$ (dashed lines) directions, and by the color map in Fig. \ref{fig.sketch}(c) as well. The potential exhibits minima at non-zero values of momentum, reaching negative values within a certain range of $k$. These minima are deeper along the $k_x$-direction, where the exciton effective mass in BP is lower. The depth and position of these minima can also be fine-tuned by adjusting the number of electrons in the 2DEG, as shown in Fig. \ref{fig.sketch}(d). Such fine control is of fundamental importance, as attractive terms in the interaction potential may result in either superfluid, supersolid, crystal or collapsed phases, depending on their magnitude.

In order to precisely distinguish these phases in the BEC density functions obtained from Eq. \eqref{GPE}, we employ the density contrast definition \cite{kirkby2023spin,smith2023supersolidity,bland2022alternating,mendoza2022exploring,blakie2020supersolidity},
\begin{equation}\label{eq. contrast}
    \mathcal{C}=\frac{\abs{\Psi}_{\text{max}}^2-\abs{\Psi}_{\text{min}}^2}{\abs{\Psi}_{\text{max}}^2+\abs{\Psi}_{\text{min}}^2},
\end{equation}
to characterize the crystallization process, and the upper bound criterion proposed by Leggett \cite{leggett1970can,leggett1998superfluid,kirkby2023spin,smith2023supersolidity,bland2022alternating,mendoza2022exploring,blakie2020supersolidity},
\begin{equation}\label{eq.leggett}
f_s=\frac{\left(\int\dd{x}\right)^2\left(\int\dd{y}\right)^2}{\int \abs{\Psi(\mathbf{r})}^2\dd{\mathbf{r}}\int \abs{\Psi(\mathbf{r})}^{-2}\dd{\mathbf{r}}},
\end{equation}
to estimate the superfluid fraction remaining in the system after the crystallization process. Here, the subscript \text{max} (\text{min}) in the former equation refers to the neighboring maximum (minimum) of the density along the $x-y$ plane, and $\int\dd{x} (\int\dd{y})$ in the latter equation denotes the length of the computational box along the $x(y)-$direction. By using these parameters, the pure superfluid phase is theoretically identified as having $\mathcal{C}=0$ and $f_s=1$, while the pure crystal phase yields $\mathcal{C}=1$ and $f_s=0$. In contrast, the supersolid phase will exhibit an intermediate value of these parameters between 0 and 1. 

In order to investigate the ground state of the excitonic BEC system, we perform numerical simulations of Eq. \eqref{GPE} using an imaginary time evolution method: a trial (random) input solution is propagated in imaginary time $t=i \tau$ over small time steps until a stable final solution is reached. This is done using the split-operator technique \cite{chaves2015split}, which allows us to avoid the integrals involved in the non-local exciton-exciton interaction potential $\Phi(\mathbf{r})$ in Eq. \eqref{GPE} by taking the potential and wave functions to reciprocal space via fast Fourier transform algorithm. All calculations are done in a 512 nm $\times$ 512 nm computational box, assuming periodic boundary conditions.

\begin{figure}[t]
\centering
\includegraphics[width = 1\linewidth]{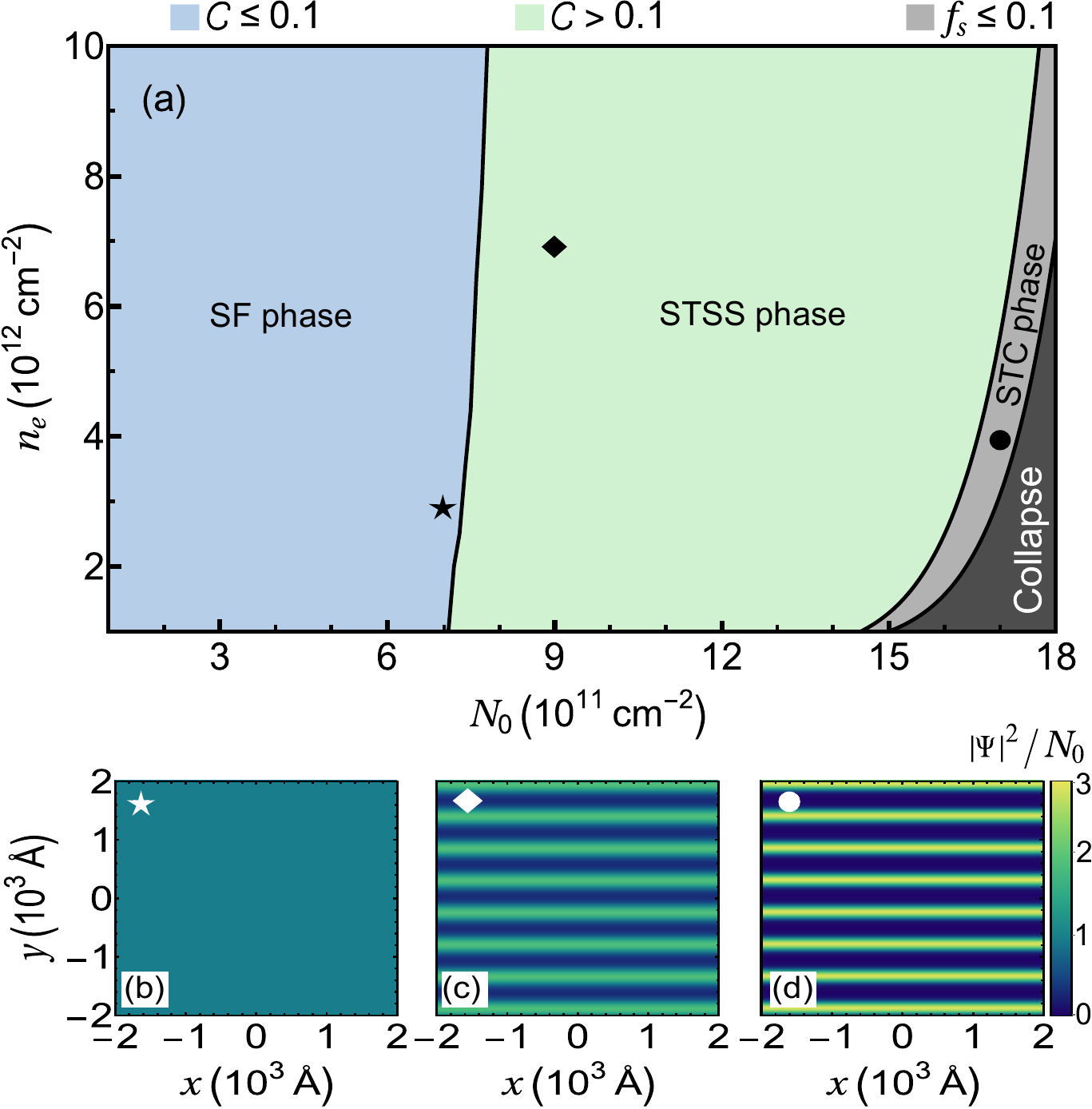}
\caption{(color online) (a) Ground state phase diagram of the excitonic BEC in a 3-layer BP slab separated by a 10-layer hBN spacer from a 2DEG in bilayer graphene, as a function of graphene doping $n_e$ and exciton density $N_0$. The color domains on the phase plots indicate distinct phases, namely superfluid (SF), striped supersolid (STSS), and striped crystal (STC), defined by their density contrast $\mathcal{C}$ and superfluid fraction $f_s$ as outlined above the phase diagram. A high density of excitons leads to a collapse in the density function. (b-d) Color plots of the condensate density $|\Psi|^2/N_0$ for the parameters marked as filled symbols in the phase diagram of panel (a).}
\label{fig.phasediagramanisotropic}
\end{figure}

\textit{Results and discussion} - To build the ground state phase diagram, we consider excitons with effective masses $m^{e}_x=0.14 m_0$, $m^e_y=1.20 m_0$, $m^h_x=0.12 m_0$, and $m^h_y=1.40 m_0$ \cite{de2017multilayered}, separated from graphene by a $N_s=10$ layer hBN spacer. As illustrated in Fig. \ref{fig.sketch}(b), this configuration exhibits a minimum in the potential $V^{\text{xx}}_{\text{eff}}$ at non-zero values of momentum, a typical scenario that can lead to crystallization in the BEC density profile \cite{matuszewski2012exciton}, by inducing a roton minimum in the excitation spectrum  \cite{plyashechnik2023coupled,cotlect2016superconductivity,kyriienko2011elementary}. Indeed, by varying the graphene doping $n_e$ and exciton $N_0$ densities, we find not only the pure superfluid phase but also a broad range of parameters where a stripe phase emerges, as shown in Fig. \ref{fig.phasediagramanisotropic}(a). Notably, adjusting the $N_0$ value reveals that these stripe phases can be classified as either a striped crystal phase or a striped supersolid phase, the latter meaning that the exciton condensate in the semiconductor plane shares crystal properties in 1D and superfluid properties in both dimensions, namely, the density function does not reach zero in the interstitial region between the stripes. We point out that first order (long-range) corrections to the contact interaction in $V^{\text{xx}}(k)$, neglected here as a first approximation, have been shown \cite{andreev2017fragmented, andreev2015autolocalization} to induce {quasi-1D} supersolid solutions under certain conditions, connected to the experimentally observed macroscopically ordered excitonic states in quantum wells \cite{butov2002macroscopically}, {as well as supersolid states in cold atom BEC}\cite{bottcher2019transient,tanzi2019observation}. However, the inclusion of this correction is a non-trivial task in {anisotropic} 2D systems, therefore, its effect on the striped phases is thus left as an exciting prospect for future works.

As previously mentioned, these phases are distinguished via Eqs. \eqref{eq. contrast} and \eqref{eq.leggett}, following the criteria typically used to characterize atomic supersolid phases \cite{kirkby2024excitations,kirkby2023spin,smith2023supersolidity,bland2022alternating,blakie2020supersolidity}. Within this framework, we identify a phase as superfluid when $\mathcal{C}\leq 0.1$ and $f_s>0.1$, while a phase with $f_s\leq0.1$ accompanied by density modulation $(\mathcal{C}>0.1)$ is categorized as a striped crystal phase. The striped supersolid phase, which lies between these regimes, is  then defined by $\mathcal{C}>0.1$ with $f_s>0.1$, as indicated in Fig.~\ref{fig.phasediagramanisotropic}(a). Differentiating the striped supersolid and striped crystal phases is important to emphasize that in the latter, although the density in the superfluid region in between the stripes never numerically reaches zero, it is still negligible, which may pose challenges to the observation of superfluid properties in this phase.

The density collapse region on the phase diagram, where the system can no longer sustain either superfluidity or stripes formation, is not adequately characterized by Eqs. \eqref{eq. contrast} and \eqref{eq.leggett}, but this phase is straightforwardly recognizable by a density function exhibiting a diverging peak with infinitesimally small width. The physics of this collapsed regime \cite{berge1998wave,adhikari2001coupled} goes beyond the mean-field description of the GPE in Eq. \eqref{GPE}, thus, within our approach, it emerges as a numerical solution rather than a physically meaningful phase. In a real sample, it is possible that other interactions beyond mean-field approximation help to stabilize the condensate and, thereby, avoid the collapse.

\begin{figure}[t]
\centering
\includegraphics[width = 1\linewidth]{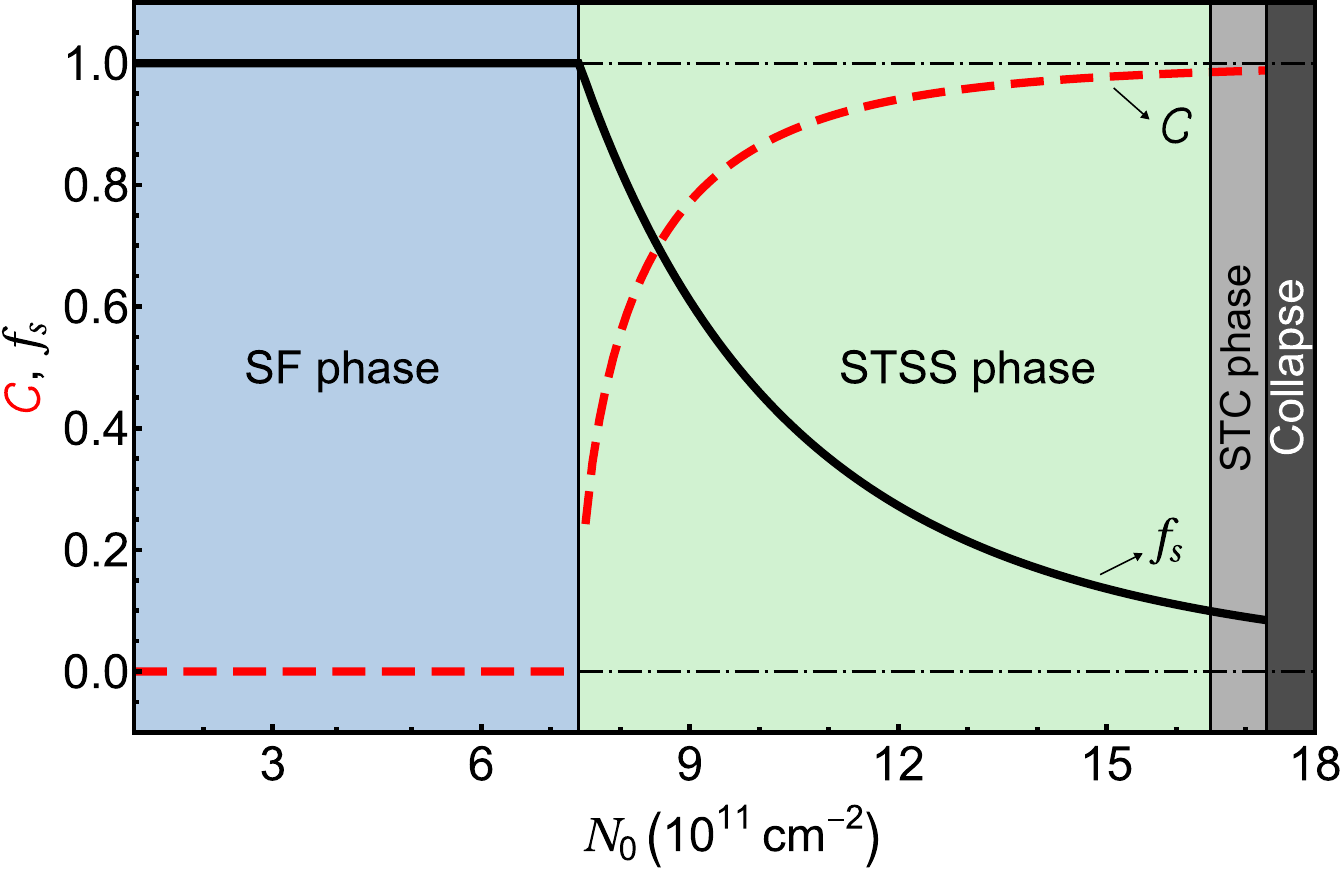}
\caption{(color online) Density contrast $\mathcal{C}$ (dashed line) and superfluid fraction $f_s$ (solid line) as a function of exciton density $N_0$ for a system with a 10-layer hBN spacer between 3-layer BP and a graphene bilayer, where a 2DEG with electron density $n_e = 4.0 \times 10^{12}$cm$^{-2}$. Horizontal dashed lines guide the eye, while the solid vertical lines demarcate the different phases, as labeled in Fig. \ref{fig.phasediagramanisotropic}(a).}
\label{fig.contrast}
\end{figure}
To provide further intuition into these phases, we exemplify spatial configurations of the superfluid, striped supersolid, and striped crystal phases in Figs. \ref{fig.phasediagramanisotropic}(b-d), respectively, corresponding to the points marked in the phase diagram in Fig. \ref{fig.phasediagramanisotropic}(a). As noted earlier, in contrast to isotropic mass semiconductors, which typically exhibit triangular or hexagonal order profiles \cite{matuszewski2012exciton,conti2023chester}, the anisotropy in BP favors a stripe order with pronounced modulation along the direction of the larger effective mass, here taken as the $y$ direction. These patterns arise spontaneously due to this anisotropy and persist over experimentally accessible densities.

In Fig. \ref{fig.contrast}, we examine the variation of the $\mathcal{C}$ and $f_s$ as a function of $N_0$ for $n_e = 4.0\times 10^{12}/\text{cm}^2$. Results show that when density modulation appears, the superfluid fraction monotonically decreases as $N_0$ increases, indicating a continuous reduction in superfluidity as density modulation strengthens. In contrast, at $N_0=7.5\times 10^{11}/\text{cm}^2$, $\mathcal{C}$ exhibits a finite jump from $\approx 0$ to $0.2$, which suggests a first-order transition, similar to what is observed in atomic supersolid systems \cite{smith2023supersolidity,poli2024excitations}. Beyond this point, both curves smoothen out, with $\mathcal{C}$ increasing more rapidly than the decrease in $f_s$, until the system reaches the striped crystal phase. Notice that even within the striped crystal phase regime, neither $f_s$ nor $\mathcal{C}$ reach their upper or lower extremes, namely, 1 and 0, respectively; instead, before convergence, the system enters the collapse regime, characterized by an abrupt increase in density accompanied by a significant negative energy shift. Collapse of a BEC of dipolar particles at high densities is a well known phenomenon in cold atoms \cite{gerton2000direct,eigen2016observation, koch2008stabilization}, originated in the interplay between attractive and repulsive contributions to particle-particle interactions. In the context of exciton-polaritons (i.e. excitons coupled to microcavity photons), collapse has been experimentally observed in AlGaAs/InGaAs quantum wells at densities similar to the ones observed here \cite{dominici2015real}, although the mechanism behind such collapse in that case is still under debate. Our results thus help to shed light on the collapse of excitons in 2D semiconductors and how to control it via 2DEG screening.

\textit{Conclusions} - Within the Gross-Pitaevskii framework, we revealed the formation of a periodic pattern in the BEC of excitons in an atomically thin BP film, where additional screening to the exciton-exciton interactions is provided by a 2DEG in a proximal layer of doped graphene. The electronic screening leads to an attractive contribution to the exciton-exciton potential that results in a minimum for this interaction at a non-zero momentum value in reciprocal space, a key ingredient for the formation of modulation in the BEC density profile. However, we demonstrate that the strong anisotropy in the exciton effective mass in BP reflects in an anisotropic potential for the screened exciton-exciton interactions, which leads to a pattern of parallel stripes in exciton density, rather than the hexagonal lattice of exciton droplets expected in isotropic condensates. Besides the pure superfluid and stripes phases, we observe a phase where these two coexist in the exciton condensate, as a striped supersolid state, which is reminiscent of a 1D supersolid state seen in alkali BECs, but without the need for any externally imposed periodicity. Our provided phase diagrams will guide further optimization efforts (doping density, exciton density and spacer thickness) to foster such 1D condensate modulation and supersolidity under realistic experimental conditions, towards premiere observation of these periodic excitonic phases in semiconducting 2D materials and heterostructures. 

\acknowledgements Discussions with A. R. P. Lima, D. Neilson, and S. Conti are gratefully acknowledged. This work was financially supported by the Brazilian Council for Research (CNPq), under the PQ, UNIVERSAL and PRONEX/FUNCAP programs, by the Brazilian Council for Higher Education (CAPES), and by the Research Foundation - Flanders (FWO-Vl).

\section{Appendix}

\subsection{Screened interaction potential}

Our system consists of a two-dimensional (2D) black phosphorus (BP) layer with $N_0$ excitons, and a graphene layer with an electron density $n_e$ that serves as a proximal 2D electron gas (2DEG). Electrons and excitons interact with each other via different electrostatic potentials: $V^{\text{xx}}_K$, $V^{\text{ee}}_k$, and $V^{\text{ex}}_k$ for the bare exciton-exciton, electron-electron, and electron-exciton interactions, respectively. However, the presence of all these charges also induce screening that modifies the bare charge-charge interactions. This screening is taken into account, in a first order approximation, as \cite{shelykh2010rotons,haug2009quantum} 
\begin{equation}
 \mathbf{V_k}^{\textbf{eff}} = \frac{\mathbf{V_k}} {1 - \mathbf{\Pi_k}\cdot\mathbf{V_k}},
\end{equation}
where 
\begin{equation}
    \mathbf{V_k} = \left(
        \begin{matrix}
             V^\text{xx}_k &  V^\text{ex}_k \\
             V^\text{ex}_k & V^\text{ee}_k
        \end{matrix} \right).
\end{equation}
and the polarization matrix is
\begin{equation}
    \mathbf{\Pi_k} = \left(
        \begin{matrix}
             \Pi^\text{x}_k &  0 \\
             0 & \Pi(k,\omega)
        \end{matrix} \right),
\end{equation}
with polarizations due to electrons and excitons given, respectively, by
\begin{equation}
\Pi(\mathbf{k},\omega) =\sum_{\mathbf{q}} \frac{f_{\mathbf{k}-\mathbf{q}}-f_{\mathbf{k}}}{\hbar\omega+i\hbar\delta+E^{e}_{\mathbf{k}-\mathbf{q}}-E^{e}_{\mathbf{k}}} 
\end{equation}
and
\begin{equation}
\Pi^\text{x}_k = \frac{2 N_0 E^\text{x}_k}{(\hbar \omega)^2 - (E^\text{x}_k)^2}.    
\end{equation}
Here, $E^{\text{x}}_k$ and $E^{\text{e}}_k$ are the energy dispersions of the excitons in the BP layer and the electrons in the 2DEG, respectively, and $f_k$ are Fermi-Dirac distributions. By performing the matrix multiplications, taking the (1,1) term of the matrix $\mathbf{V_k}^{\textbf{eff}}$ and substituting the bare charge-charge potentials given by Eqs. (4) and (6) of the main manuscript, one obtains $V^{\text{xx}}_{\text{eff}}$ as in Eq. (2) of the main text.

Electron-electron interactions $V_{k}^{\text{ee}}$ are assumed to be in the Coulomb form with the dielectric constant of the environment, as it is usual for graphene, due to its thin layer thickness. Exciton-exciton interactions $V_{k}^{\text{xx}}$ for electrons and holes spatially separated by a distance $d$ along the out-of-plane direction could be taken as dipolar interactions $V_{dd}(r) = e^2\big /4 \pi \epsilon(2/r-2/\sqrt{r^2+d^2})$ as a first approximation. For a constant exciton density $N$, the energy blueshift as one increases the density reads  
\begin{equation}
\Delta E =\int{NV_d(r)d^2r} = \frac{e^2d}{\epsilon}N    
\end{equation}
which allows one to safely replace the dipolar interaction in a Gross-Pitaevskii formalism for an exciton superfluid by a contact interaction $V_{dd}(r-r') \sim e^2\big/2\pi\epsilon\delta(r-r')$, so that
\begin{eqnarray}
\mu \approx \frac{1}{N} \int \left[\int{V_{dd}(r-r') {|\psi(r')|}^2d^2r'}\right] {|\psi(r)|}^2 d^2r = \nonumber\\
 \frac{1}{N} \int \frac{e^2d}{\epsilon} {|\psi(r)|}^4 rdr = \frac{e^2d}{\epsilon}N = \Delta E.
\end{eqnarray}
This procedure, known as the ``capacitor model'' for excitonic interactions, successfully describes the experimentally observed blueshift in a qualitative way. However, the actual slope of the blueshift as a function of $N$ in experiments is much lower than predicted by the capacitor model, due to correlations in the exciton gas \cite{steinhoff2024exciton}, which calls for a correction factor $\xi$ in the interaction, such that $V^{\text{xx}}(r-r') = \frac{e^2d \xi}{2\pi\epsilon}\delta(r-r')$. The Hankel transform of this potential is the expression for $V_k^{\text{xx}}$ given in Eq. (6) of the main text. Electron-exciton interactions $V_{k}^{\text{ex}}$ are the discussed in what follows.

\subsection{Electron-exciton Interaction potential}

We briefly comment on the calculation of the electron-exciton interaction potential given by Eq. (4) in the main text. For an electron in the 2DEG with momentum $\mathbf{k'}$ and an exciton with center-of-mass momentum $\mathbf{K}$, this involves calculating the exchange term:
\begin{equation} \label{eq1}
    V^{\text{ex}}_{\mathbf{k}}=\sum_{\mathbf{k'},\mathbf{K}}V_\mathbf{k}a_{\mathbf{K}+\mathbf{k}}^\dagger b_{\mathbf{k}'-\mathbf{k}}^\dagger a_\mathbf{K} b_\mathbf{k'}
\end{equation}
where $a_\mathbf{K}$ and $b_\mathbf{k'}$ denote the creation operators (with corresponding annihilation operators $a_\mathbf{K}^\dagger$ and $b_\mathbf{k'}^\dagger$)  for an exciton in the BP slab and an electron in the 2DEG, respectively. The momentum $\mathbf{k}$ corresponds to the momentum exchanged due to the Coulomb interaction $V_{\mathbf{k}}$, defined, in real space, as:
\begin{equation}\label{eq.coulomb}
V_{\text{C}}(\mathbf{r}',\mathbf{r_e},\mathbf{r_h}) = \frac{e^2}{4 \pi \epsilon}\left(\frac{1}{\abs{\mathbf{r'}-\mathbf{r}_h}}-\frac{1}{\abs{\mathbf{r'}-\mathbf{r}_e}}\right),
\end{equation}
where $\mathbf{r'}$ refers to the position of the electron in the 2DEG, and $\mathbf{r}_{e}$ and $\mathbf{r}_h$ denote the positions of the electron and hole that make up the exciton.

To evaluate Eq. \eqref{eq1} under this potential, we assume that the exciton wave function takes the form $\Psi_{\text{x}}(\mathbf{r})=\mathcal{N} \delta(z_e-d/2)\delta(z_h+d/2)e^{i\mathbf{K}\cdot \mathbf{R}}e^{-\sqrt{(x \mu_x)^2+(y \mu_y)^2}/a_B}$, where  $\mathcal{N}=\sqrt{4/\pi a_B^2}$ is the normalization factor,  $\mathbf{r}$ is the relative exciton coordinate, and  $\mathbf{R}=\left[(m_x^e x^e + m_x^e x^h)/m_x^{\text{x}}, (m_y^e y^e + m_y^h y^h)/m_y^{\text{x}}\right]$ denotes the exciton's center-of-mass position, with $m_{x,y}^{e,h}$ and $m_{x,y}^{\text{x}}=m_{x,y}^{e}+m_{x,y}^{h}$ being the effective masses of the electron $e$, hole $h$, and exciton $X$, respectively, along the direction $\mathbf{r}$. The anisotropy in the wave function for the relative coordinate of the exciton is taken into account by assuming an effective direction dependent Bohr radius $a_B/\mu_{x,y}$, where $\mu_{x,y} = (1/m^e_{x,y} + 1/m^h_{x,y})^{-1}$ is the electron-hole reduced mass.

Under theses considerations, Eq. \eqref{eq1} can be conveniently evaluated over real space, yielding:
\begin{widetext}
\begin{equation}
    \begin{split}
         V^{\text{ex}}(\mathbf{k})&=\frac{e^2 \mathcal{N}^2}{4\pi \epsilon}\int \left[\frac{e^{-i\mathbf{k}\cdot \mathbf{R}}e^{i\mathbf{k}\cdot \mathbf{r'}} e^{-2r/a_B}}{\sqrt{\left(\mathbf{r'}-\mathbf{R}-\beta^h\mathbf{r}\right)^2+(L-d/2)^2}}-\frac{e^{-i\mathbf{k}\cdot \mathbf{R}}e^{i\mathbf{k}\cdot \mathbf{r'}} e^{-2r/a_B}}{\sqrt{\left(\mathbf{r'}-\mathbf{R}+\beta^e\mathbf{r}\right)^2+(L+d/2)^2}}\right]\dd{\mathbf{r}}\dd{\mathbf{R}}\dd{\mathbf{r'}}.
    \end{split}
\end{equation}
\end{widetext}
Here, $\beta_{x,y}^{e,h}=m_{x,y}^{e,h}/m_{x,y}^{\text{x}}$, $\beta^{e,h}\mathbf{r} = \beta_x^{e,h}x \mathbf{\hat x}+\beta_y^{e,h}y \mathbf{\hat y}$, $L$ represents the out-of-plane distance between the 2DEG and the center of the BP slab, which appears in this expression as a result of the integration over $z$, and $d$ is the vertical separation between electrons and holes in the slab hosting excitons, with $z=d/2$ and $z=-d/2$ being the out-of-plane localization of the electron and hole, respectively. 

After further calculation, we arrive at the more compact form:
\begin{widetext}
\begin{equation}\label{eq2}
    V^{\text{ex}}(\mathbf{k})=\frac{e^2 \mathcal{N}^2}{4\pi \epsilon}\int \left[\frac{1}{\beta_x^h\beta_y^h} (f_h * g_h)\left(\mathbf{r'}-\mathbf{R}\right)-\frac{1}{\beta_x^e\beta_y^e} (f_e * g_e)\left(\mathbf{R}-\mathbf{r'}\right)\right]e^{-i\mathbf{k}\cdot (\mathbf{R}-\mathbf{r'})}\dd{\mathbf{R}}\dd{\mathbf{r'}}.
\end{equation}
\end{widetext}
This expression shows that $V^{\text{ex}}(\mathbf{k})$ is given by the Hankel transform of a convolution, $(f_{e,h}*g_{e,h})(\mathbf{x})$, which, in turn, is equivalent to the product of the Hankel transforms of the functions involved. Here, $f_{e,h}(\beta_{x,y}^{e,h}\mathbf{r})=\left[\left(\mathbf{x}-\beta^{e(h)}\mathbf{r}\right)^2+(L\pm d/2)^2\right]^{-1/2}$, where, as in the main text, the $-(+)$ sign refers to the $e(h)$ component, and $g_{e,h}(\beta^{e,h}\mathbf{r})=\exp\left(-2\beta^{e,h}\sqrt{(x \mu_x)^2+(y \mu_y)^2}/{\beta^{e,h} a_B}\right)$, thus leading to the final form:
\begin{widetext}
\begin{equation}
     V^{\text{ex}}(\mathbf{k})=\frac{e^2}{2\,\epsilon\,k}\left\{\frac{e^{-k\left(L-d/2\right)}}{\left[1+\left(\frac{\beta_x^h a_B k_x}{2\mu_{x}}\right)^2+\left(\frac{\beta_y^h a_B k_y}{2\mu_{y}}\right)^2\right]^{3/2}}-\frac{e^{-k\left(L+d/2\right)}}{\left[1+\left(\frac{\beta_x^e a_B k_x}{2\mu_{x}}\right)^2+\left(\frac{\beta_y^e a_B k_y}{2\mu_{y}}\right)^2\right]^{3/2}}\right\}.
\end{equation}
\end{widetext}
It is straightforward to verify that for the isotropic mass case $m_x^{e,h} = m_y^{e,h}$, this expression agrees with those in previous papers \cite{plyashechnik2023coupled, matuszewski2012exciton}. 

For a more accurate account of this interaction, further improvement could be made: the Coulomb form Eq. \eqref{eq.coulomb} is an approximation in our system, since it has a dielectric constant $\epsilon(z)$ that changes along the out-of-plane direction $z$, as it has different values in the BP, hBN and graphene regions. In this case, obtaining the correct interaction requires solving a Poisson equation $\nabla \cdot [\epsilon(z)\nabla \phi] = q\delta(\mathbf{r} - \mathbf{r}_c) $. For particles in the same layer, assuming $kw \ll 1$, where $w$ is the layer width, and for a surrounding medium with dielectric constant much lower than that of the layer, N. Rytova \cite{rytova2018screened} and L. Keldysh \cite{keldysh1979coulomb} demonstrated that the effective interaction potential still exhibit the form of a Coulomb potential in reciprocal space, but with a $k$-dependent static dielectric constant $\epsilon(k)$ that can be approximated as $\epsilon(k) \approx \epsilon_s(1+r_0 k)$, where $\epsilon_s$ is the dielectric constant of the surrounding material, and $r_0$ is a screening length. This linear approximation, around $k \approx 0$, could be relevant for $V^{\text{ee}}$ and $V^{\text{xx}}$, where interacting charges are in the same layer. However, in the former, due to the small width of graphene, it is customary \cite{polini2008plasmons} to assume a Coulomb interaction with the dielectric constant of the surrounding medium (i.e. $r_0 k \approx 0$) as a good approximation, as we did in Eq. (6) of the manuscript. Indeed, we have checked that using an estimate for the dielectric constant of graphene \cite{annies2023intrinsic}, $r_0 = w \epsilon_{graph}/2\epsilon \approx 5$\AA\,, so that $k r_0 \ll 1$ for the range of $k$ in our calculations. In $V^{\text{xx}}$, it is not clear how such correction in $\epsilon(k)$ around $k \approx 0$ would affect the contact interaction we assumed in the manuscript. Calculating the actual exciton-exciton interaction in 2D materials is a difficult task; what is clear, though, is the fact that the classical dipole-dipole interactions \cite{lee2009exciton} and even the capacitor model approximation \cite{lobanov2016theory, erkensten2021exciton, erkensten2023electrically} are strongly suppressed by exchange and correlations \cite{steinhoff2024exciton}, which explains the phenomenological $\xi$ factor used in Eq. (6) of the main text, but the actual form of this potential is left as an interesting topic for future investigation. As for $V^{\text{ex}}$, the Rytova-Keldysh form does not apply, since the charges are in separate layers. A thorough study for intra- and inter-layer interactions beyond Rytova-Keldysh approximations is presented in Ref \cite{cavalcante2018electrostatics}, where it was demonstrated that for charges separated by 10 layers of hBN, as it is the case here, the Coulomb interaction is still a good approximation. Besides these electrostatic screening corrections, vertex corrections to the random phase approximation used here could also be added, but they are beyond our current mean field model and may not have a significant effect in our system, although further investigation is recommended to confirm this assumption \cite{plyashechnik2023coupled}. 

\subsection{Roton in the excitation spectrum}

\begin{figure}[]
\centering
\includegraphics[width=0.8\linewidth]{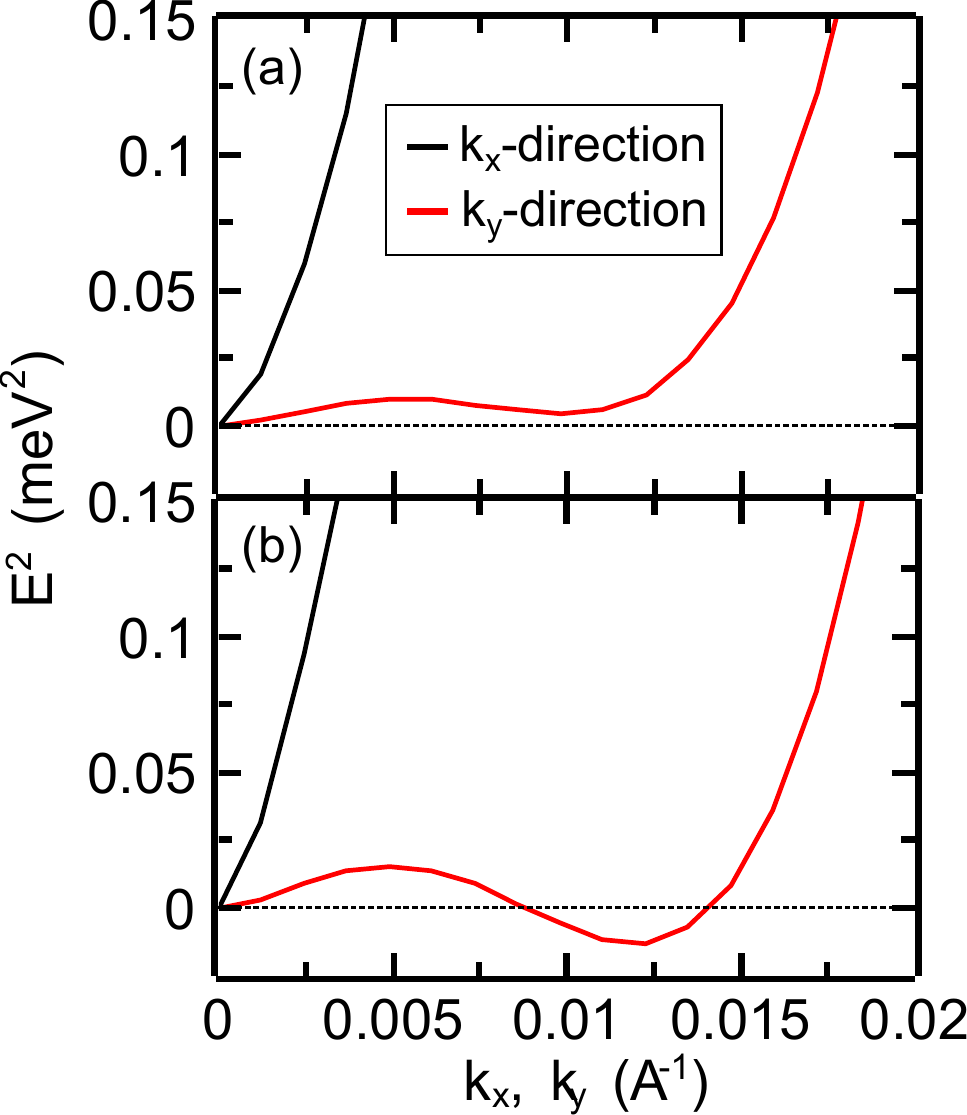}
\caption{(color online) Excitation spectra of the excitonic condensate along $k_x$ (black) and $k_y$ (red) directions, for the same system as in Figs. 2 and 3 of the main manuscript, considering $n_e = 3\times10^{12}$cm$^{-2}$ and $N_0 =$ (a) $6 \times 10^{11}$cm$^{-2}$ and (b) $1 \times 10^{12}$cm$^{-2}$.}
\label{fig.roton}
\end{figure}

Figure \ref{fig.roton} shows examples of excitation spectra obtained by $E^2(k_x,k_y)=\left(E^{\text{x}}_\mathbf{k}\right)^2+2N_0 V^{\text{xx}}_{eff}E^{\text{x}}_\mathbf{k}$, with the screened exciton-exciton interaction $V^{\text{xx}}_{eff}$ obtained as explained in the first Subsection of this Appendix, for the same system as in Figs. 2 and 3 of the main manuscript, considering $n_e = 3\times10^{12}$cm$^{-2}$ and $N_0 =$ (a) $6 \times 10^{11}$cm$^{-2}$ and (b) $1 \times 10^{12}$cm$^{-2}$. Due to the mass anisotropy, the excitation spectrum is strongly anisotropic and the roton minimum is visible only along the $k_y$-direction, namely, the direction with larger effective mass. Also, the roton minimum crosses $E = 0$ and becomes the lowest energy state, as illustrated in Fig. \ref{fig.roton}(b), only for $N_0 \gtrsim 7 \times 10^{11}$cm$^{-2}$, i.e. after crossing the line that separates superfluid and striped supersolid phases in Fig. 2 of the main manuscript. Consequently, the exciton condensate develops a density modulation along the $y$-direction only for parameters in the striped supersolid and striped crystal sectors of that phase diagram.

\subsection{Time evolution method}

As mentioned in the main text, the ground state of the excitonic condensate is obtained by imaginary time evolution of a random initial wave function. The time evolution is performed here by means of the so-called split-operator technique \cite{chaves2015split}, where one can deal with kinetic and potential energy operators separately, each in its more convenient (real or reciprocal) space. This is particularly important for the Gross-Pitaevskii equation describing dipolar gases, where the calculation of the non-local potential term $\Phi(\mathbf{r},t) = \int V^{\text{xx}}_{\text{eff}}(|\mathbf{r}-\mathbf{r'}|,\omega)\abs{\Psi(\mathbf{r'},t)}^2 d\mathbf{r'}$ is quite time consuming, as it involves an integration over the entire space that needs to be performed at each time step. This is partially mitigated by using a fast Fourier transform (FFT) algorithm to take functions to reciprocal space, where this convolution integral is rewritten as a simple multiplication $V^{\text{xx}}_{\text{eff}}(\mathbf{k},\omega)\abs{\Psi(\mathbf{k},t)}^2$.

In the split-operator technique, the time evolution is approximated as
\begin{widetext}
\begin{equation}\label{eq.SMsplitoperator0}
  \Psi(\mathbf{r},t+\delta t) \approx \exp[-\frac{i}{2\hbar}\hat{\Phi}(\mathbf{r},t)\delta t]\exp[-\frac{i}{\hbar}\hat{T}\delta t]\exp[-\frac{i}{2\hbar}\hat{\Phi}(\mathbf{r},t)\delta t]\Psi(\mathbf{r},t). 
\end{equation}
\end{widetext}
In this approach, two approximations were taken: (i) $ \int_{t}^{t+\delta t}\Phi(\mathbf{r},t)\delta t \approx \Phi(\mathbf{r},t)\delta t$, which is rooted in the mean value theorem for small $\delta t$ $(\sim 10^{-14} s)$, and (ii) $\exp{A+B} \approx \exp{A/2}\exp{B}\exp{A/2}$, which originates in the Suzuki-Trotter expansion of the exponential and includes an error of the order of $O(\delta^3)$. Both approximations are justified given the small $\delta t$ time-steps considered here.

In order to obtain $\Psi(\mathbf{r},t+\delta t)$ with Eq. \eqref{eq.SMsplitoperator0}, we first employ a FFT to take $\Psi(\mathbf{r},t)$ and the argument of the exponential to reciprocal space, where the multiplication is performed as 
\begin{widetext}
\begin{equation}\label{eq.SMsplitoperator}
  \Psi(\mathbf{k},t+\delta t) = \exp[-\frac{i}{2\hbar}V^{\text{xx}}_{\text{eff}}(\mathbf{k})\abs{\Psi(\mathbf{k},t)}^2\delta t]\exp[-\frac{i}{\hbar}\left(\frac{\hbar^2k_x^2}{2 m_x^{\text{x}}}+\frac{\hbar^2k_y^2}{2 m_y^{\text{x}}}\right)\delta t]\exp[-\frac{i}{2\hbar}V^{\text{xx}}_{\text{eff}}(\mathbf{k})\abs{\Psi(\mathbf{k},t)}^2\delta t]\Psi(\mathbf{k},t). 
\end{equation}
\end{widetext}
By the end of the procedure, we employ an inverse FFT to take $\Psi(\mathbf{k},t+\delta t)$ back to real space, which allows us to plot color maps of the condensate density as shown in Fig. 2 of the main text.  

Notice that the frequency dependence of the $V^{\text{xx}}_{\text{eff}}$ potential can lead to retarded interaction, which complicates the time evaluation of this term. To avoid this, as in Ref. \cite{matuszewski2012exciton}, we average the potential over a frequency range limited by the exciton dissociation frequency $\omega_d=E_B/\hbar$, which in turn leads to a non-retarded GPE, with a $\omega$-independent $\Phi(\mathbf{r},t)$, as in Eq. \eqref{eq.SMsplitoperator}.

\subsection{Isotropic mass case}

\begin{figure}[]
\centering
\includegraphics[width=0.8 \linewidth]{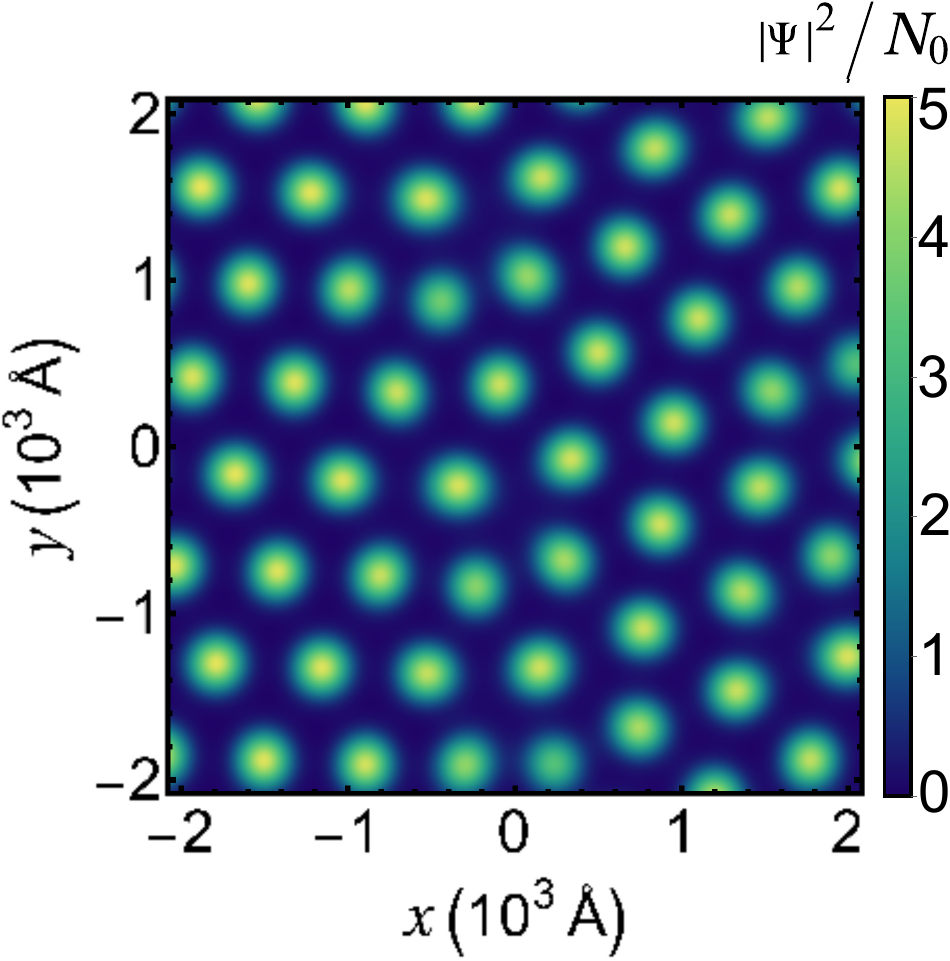}
\caption{(color online) Colormap of the particle density for the case of a material with isotropic masses for electrons and holes in the exciton, $m_e^x = m_e^y = 1.20 m_0$ and $m_h^x = m_h^y = 1.40 m_0$. The exciton layer is separated from the bilayer graphene 2DEG by a $N_s = 11$ layers hBN spacer. The exciton density is $N_0 = 1.9 \times 10^{12} cm^{-2}$ and the electron density is $n_e = 3.0 \times 10^{12} cm^{-2}$. All other parameters are kept the same as in the main text.}
\label{fig.isocrystal}
\end{figure}

As a proof-of-concept, we look for a situation where instead of observing stripes as the density modulated phase, we observe a crystal structure in the form of a triangular lattice, as reported in Ref. \cite{matuszewski2012exciton}. Indeed, Fig. \ref{fig.isocrystal} illustrates a crystalized phase observed for system where electrons and holes have isotropic effective mass with the same value as the one of the heaviest mass direction in BP, namely, $m_e^x = m_e^y = 1.20 m_0$ and $m_h^x = m_h^y = 1.40 m_0$. In this case, in order to obtain the crystalized phase, we also needed to weaken the interaction potential, therefore, we added one more hBN layer to the spacer between the 2D semiconductor and the bilayer graphene 2DEG, i.e., we assume a $N_s = 11$ layers spacer. All other parameters are kept the same in this calculation. This result helps validating our model, by comparing to results already established in the literature \cite{matuszewski2012exciton}, while confirming that the stripe structure is a consequence of mass anisotropy. 

\bibliography{references} 

\end{document}